\documentclass[twocolumn,superscriptaddress,showpacs, longbibliography, aps, prb]{revtex4-1}
\usepackage{color}
\usepackage{hyperref}
\usepackage{amsmath, amssymb}
\hypersetup{colorlinks=true,breaklinks,linkcolor=blue,urlcolor=blue,citecolor=blue}
\usepackage{graphicx}% Include figure files
\usepackage{dcolumn}% Align table columns on decimal point
\usepackage{bm}% bold math
\usepackage{xspace}

\def\vec#1{\boldsymbol #1}

\usepackage{natbib}

\begin{document}

\preprint{}

\title{$Ab$ $initio$ derivation and exact-diagonalization analysis of \\
low-energy effective Hamiltonians for $\beta^\prime$-X[Pd(dmit)$_2$]$_2$}

\author{Kazuyoshi Yoshimi}
\affiliation{Institute for Solid State Physics,~University of Tokyo,~5-1-5 Kashiwanoha, Kashiwa, Chiba 277-8581, Japan}
\author{Takao Tsumuraya}
\affiliation{Priority Organization for Innovation and Excellence, Kumamoto University, 2-39-1 Kurokami, Kumamoto 860-8555, Japan}
\author{Takahiro Misawa}
\affiliation{Beijing Academy of Quantum Information Sciences, Haidian District, Beijing 100193, China}

\date{\today}% It is always \today, today,
             %  but any date may be explicitly specified

\begin{abstract}
The molecular solids $\beta^\prime$-$X$[Pd(dmit)$_2$]$_2$ (where $X$ represents a cation) are
typical compounds whose electronic structures are described by single-orbital Hubbard-type Hamiltonians with geometrical frustration.
Using the $ab$ $initio$ downfolding method,
we derive the low-energy effective Hamiltonians for $\beta^\prime$-$X$[Pd(dmit)$_2$]$_2$ with
available room- and low-temperature structures.
We find that the amplitudes of the Coulomb interactions 
and the anisotropy of the hopping parameters in the effective Hamiltonians are sensitive 
to the changes in the lattice constants induced by lowering the temperature. 
The obtained effective Hamiltonians are analyzed 
using the exact diagonalization method with the boundary condition average.
We find that a significant reduction of the antiferromagnetic
ordered moment occurs in the effective Hamiltonian of $\beta^\prime$-EtMe$_3$Sb[Pd(dmit)$_2$]$_2$ 
with the low-temperature structure. The reduction is consistent with the
quantum spin liquid behavior observed in experiments.
The comprehensive derivations of the effective Hamiltonians and exact-diagonalization
analyses of them will clarify 
the microscopic origins of the exotic quantum states of matter found 
in $\beta^\prime$-$X$[Pd(dmit)$_2$]$_2$ such as the quantum spin liquid behavior. 
\end{abstract}

\maketitle

\section{Introduction}
The single-orbital Hubbard model and its extensions have been studied 
as canonical models that show exotic quantum phases of matter
such as high-$T_{\rm c}$ superconducting phases~\cite{Imada_RMP1998} 
and quantum spin liquids (QSLs)~\cite{Balents_Nature2010}.
It has been shown that some families of molecular solids 
such as (BEDT-TTF)$_2X$ with $\kappa$ and $\beta^{\prime}$-type dimer 
arrangements offer an ideal platform for realizing 
the single-orbital Hubbard model~\cite{Ishiguro, Kanoda2006, Koretsune_PRB2014}.
Although most molecular solids have a complex structure typically containing more than 100 atoms  
in the unit cell, the low-energy band structures near the Fermi level are simple, 
i.e., one frontier molecular orbital of the constituent molecule 
contributes to the formation of the conduction bands and the other orbitals are sufficiently far from the Fermi level~\cite{Kanoda_Kato_2011ARCMP}. 
In fact, the simple band structures near the Fermi level have been confirmed by several experiments, such as de Haas-van Alphen (dHvA) measurements~\cite{Uji_1997}. 
The simple isolated band structures 
in the molecular solids are in sharp contrast with those of 
transition metal oxides such as cuprates, 
where $p$ orbitals in oxygen often largely hybridize 
with $d$ orbitals in transition metals~\cite{Hybertsen_PRB_1990,Hirayama_PRB2018}.

Among molecular solids, the anion radical salt of $\beta^\prime$-$X$[Pd(dmit)$_2$]$_2$ (dmit = 1,3-dithiole-2-thione-4,5-dithiolate; $X$ = a monovalent closed-shell cation) is 
a typical compound whose electronic structures can be described by the single-orbital Hubbard-type Hamiltonian. 
In these solids, the [Pd(dmit)$_2$] unit is strongly dimerized and the dimers form an anisotropic triangular lattice in each anion layer. 
In the dimer unit, both the HOMO and LUMO of the Pd(dmit)$_2$ monomer form bonding and antibonding pairs, and 
a half-filled band is mainly formed through the antibonding HOMO pair~\cite{Canadell_dmit_1989, Canadell_SSC1990, Kato_Chem_Rev, Miyazaki99_dmit}.
Experimentally, by changing the type of cation $X$,
a rich variety of ground states such as antiferromagnetically (AF) ordered states~\cite{Tamura_JPhys2002}, 
QSL states~\cite{Itou_dmit_PRB2008, Yamashita2010, Yamashita2011}, 
and charge ordered (CO) states~\cite{Nakao_JPSJ_Et2Me2Sb, Tamura_CPL2004} have been found.

In our previous study~\cite{Misawa_2020_PRRes}, we have shown that the single-orbital extended Hubbard-type Hamiltonians obtained in $ab$ $initio$ calculations successfully reproduce the cation dependence of the magnetic properties in the dmit salts, 
i.e., the peak value of the spin structure factors is significantly suppressed around the QSL compound X $=$ EtMe$_3$Sb salt (Et = C$_2$H$_5$ and Me = CH$_3$, hereafter called the X salt, for simplicity).
We found that the magnetic moment of EtMe$_3$Sb salt reduces
only in the effective Hamiltonian of the low-temperature (LT) structure of 4~K 
and not in that of the room temperature (RT) structure.
This result indicates that changes in a lattice structure (i.e., the distance between dimers) due to a decrease in temperature significantly influence the electronic structures of the dmit salts.

To date, changes in the lattice structures for decreasing temperatures have not been seriously
examined in the context of {\it ab initio} derivations of the low-energy effective Hamiltonians.
Most studies of $ab$ $initio$ derivations of the low-energy effective Hamiltonians have employed the experimental structure at RT to avoid the effects of symmetry breaking at LT, 
such as magnetic orderings~\cite{Crystals12_Kato, Powell_PRL12,Jacko_PRB2013, Tsumu_Pd_dmit2_13, Seo_2015JPSJ, 1D_Hab_Valenti}.
This treatment is justified when the changes in the lattice constants are small.
However, molecular solids are flexible 
because they are loosely bounded by van der Waals interactions.
Thus it is expected that changes in the lattice structures by lowering the temperature significantly affect the electronic structures of the dmit salts.
Although it was reported that the transfer integrals between dimers 
for LT structures of EtMe$_3$Sb (4.5~K)~\cite{Kato_unpub} and Me$_4$P salts (8~K)~\cite{Kato_Me4P_LT} show a larger anisotropy of the triangular lattice than those at RT~\cite{Nakamura, Tsumu_Pd_dmit2_13, Misawa_2020_PRRes},
the effects of lowering the temperature for other compounds has not been clarified.
More recently, the LT structures of Et$_2$Me$_2$As, Me$_4$As, and Me$_4$Sb salts have been reported~\cite{KUeda_EtMe3Sb2018}. 
Using these LT structures, it is now possible to clarify how the changes in the structures affect the microscopic parameters and physical quantities in the effective Hamiltonians of the dmit salts.

In this paper, using the $ab$ $initio$ downfolding method~\cite{Aryasetiawan_PRB2004,Imada_JPSJ2010},  
we derive low-energy effective Hamiltonians 
for nine dmit salts with RT structures
and five dmit salts with LT structures.
We clarify how the changes in the lattice constants affect the microscopic parameters in 
the effective Hamiltonians, such as the transfer integrals and Coulomb interactions.
By solving the derived effective Hamiltonians using the exact diagonalization method,
we also clarify the effects of changes in the lattice constants on the physical quantities, such as the spin structure factors and the charge gap. 
We find that all the Hamiltonians for the
LT structures show stripe-type magnetic ordered correlations and significant reductions in the magnetic ordered moment occur around the EtMe$_3$Sb salt.
This result is consistent with the compound dependence of the N\'{e}el temperatures
and the quantum spin liquid behavior observed in EtMe$_3$Sb salt~\cite{Itou_dmit_PRB2008, Yamashita2010, Yamashita2011}.

The remainder of this paper is organized as follows. 
In Sec. \ref{Sec:Methods}, 
the methods to derive low-energy effective Hamiltonians and analyze them are described. 
In Sec. \ref{Sec:NumericalResults}, 
the values of parameters such as transfer integrals and Coulomb interactions, 
which constitute the effective Hamiltonians, are listed for each material and temperature, 
and their trends are discussed.
In addition, by analyzing these models using the exact diagonalization method, 
we obtain the compound dependence of the spin structure factors and the charge gap.
In the exact-diagonalization analyses,
to reduce the finite-size effects, 
we performed the boundary-condition average~\cite{twisted_boundary}.
We also discuss how our obtained results can be observed in experiments.
Section \ref{Sec:Summary} contains a summary of the paper.

\section{Methods}\label{Sec:Methods}
\subsection{Derivation of low-energy effective Hamiltonians}
In this study, the low-energy effective Hamiltonians are derived from the nonmagnetic band structure obtained using first-principles density functional theory (DFT) calculations~\cite{H-K_1964, Kohn_Sham}. In the series of DFT calculations, we refer to the experimental structures at LT and RT~\cite{Crystals12_Kato, KUeda_EtMe3Sb2018}.
The RT structures of the nine members of $\beta^\prime$ salts are isostructural, 
with a face-centered monoclinic structure of the space group $C$2/$c$. 
No structural phase transition is observed by lowering the temperature 
for all the salts except for the Et$_2$Me$_2$Sb salt, which causes CO.
In this work, we study the cation dependence of the electronic state of the lowest temperature for EtMe$_3$Sb (5~K), Me$_4$P (8~K), Me$_4$As (5~K), Me$_4$Sb (5~K), and Et$_2$Me$_2$As (5~K) salts for which the experimental structures have been reported~\cite{Kato_Me4P_LT, KUeda_EtMe3Sb2018}.

The preset first-principles calculations were performed using the pseudopotential method based on the optimized norm-conserving Vanderbilt (ONCV) formalism with plane-wave basis sets~\cite{Hamann_ONCV2013, Schlipf_CPC2015}, which is implemented in \texttt{Quantum Espresso (version 6.3)}~\cite{QE}.  
The exchange-correlation functional of the generalized gradient approximation (GGA) proposed by Perdew, Burke, and Ernzerhof (PBE) was employed~\cite{GGA_PBE}. 
The cutoff energies for plane waves and charge densities are 70 and 280 Ry, respectively. 
A $5\times5\times3$ uniform $\bm{k}$-point mesh was used for all the ${\beta^\prime}$ salts with a Gaussian smearing method during self-consistent loops. 
All the hydrogen positions were relaxed through the geometrical optimization procedure 
of the first-principles method because it is difficult to accurately determine C--H bond distances from x-ray diffraction measurements.

\begin{figure*}[t] 
\begin{center} 
\includegraphics[width=0.7 \textwidth]{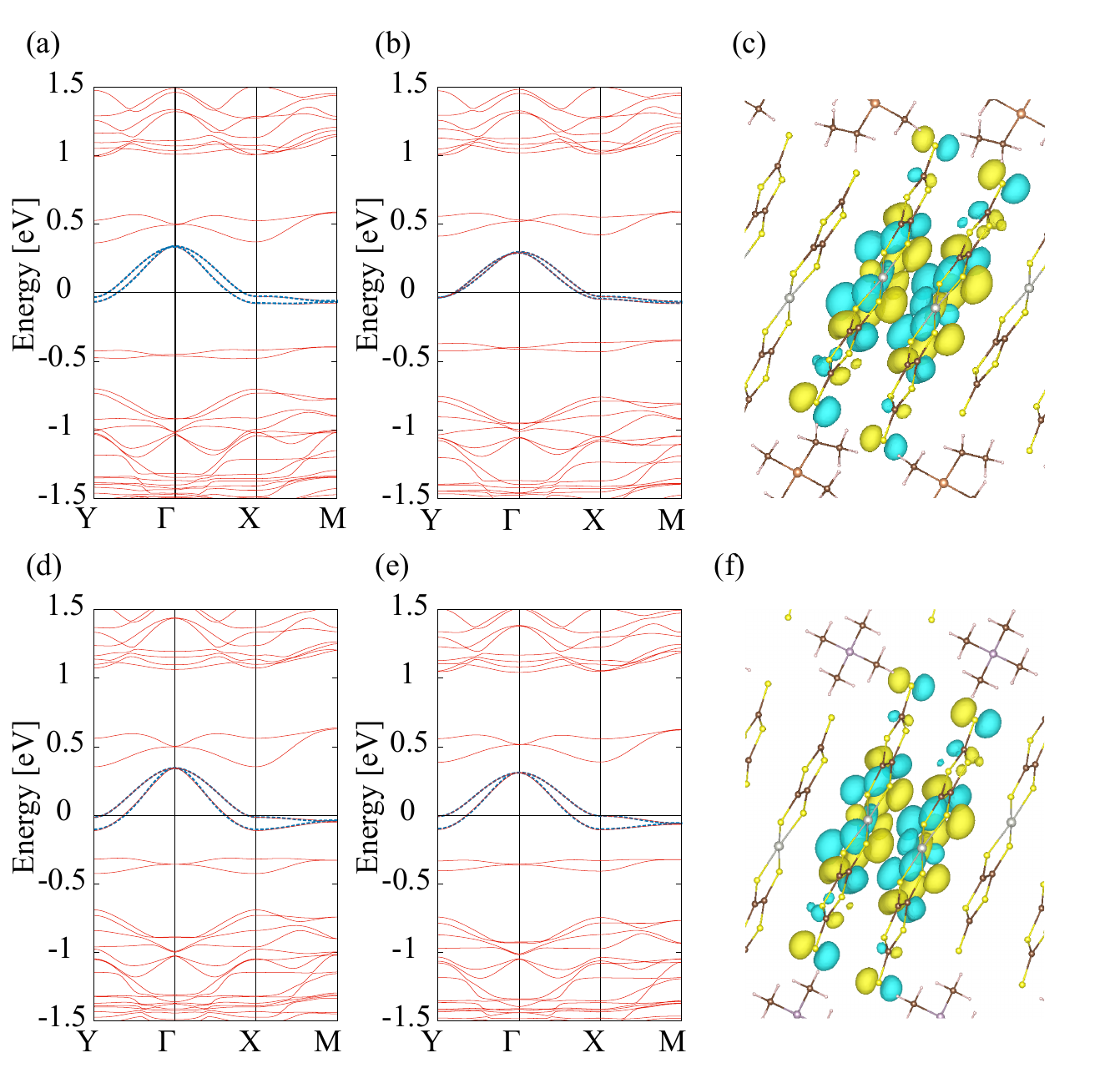}
\vspace{-0.5cm} 
\caption{Band dispersion of $\beta^\prime$-EtMe$_3$Sb[Pd(dmit)$_2$]$_2$ at (a) $5$~K and (b) RT. 
The solid lines are obtained by DFT calculations, while the broken lines are obtained using the maximally localized Wannier functions (MLWFs). 
The energy is shifted so that the Fermi energy is zero. 
(c) MLWF of $\beta^\prime$-EtMe$_3$Sb[Pd(dmit)$_2$]$_2$ at $5$~K. 
Band dispersion of $\beta^\prime$-Me$_4$P[Pd(dmit)$_2$]$_2$ at (d) $8$~K and (e) RT, 
and (f) MLWF of $\beta^\prime$-Me$_4$P[Pd(dmit)$_2$]$_2$ at $8$~K.
MLWFs are drawn by using \texttt{VESTA}~\cite{VESTA}.} 
\label{fig-bands}
\end{center}
\end{figure*}

Using the obtained Bloch functions, we generated maximally localized Wannier functions (MLWF) for the half-filled bands crossing the Fermi level using \texttt{RESPACK}~\cite{RESPACK}.
Then, the transfer integral between MLWFs at $\vec{R}_1$ and $\vec{R}_2$ sites is given as follows:
\begin{align}
t_{\vec{R}_1\vec{R}_2}=
\langle\phi_{\vec{R}_1}| H_0|\phi_{\vec{R}_2}\rangle. \label{eq:transfer}
\end{align}
Here, $|\phi_{\vec{R}}\rangle = c_{\vec{R}}^{\dagger}|0\rangle$, where  $c_{\vec{R}}^{\dagger}$ is the creation operator of an electron in the Wannier orbital at ${\vec{R}}$ site and $H_0$ represents the non interaction term of the $ab$ $initio$ Hamiltonians.
In Fig.~\ref{fig-bands}, we show the band structure and the MLWFs of EtMe$_3$Sb and Me$_4$P salts, which we will discuss in Sec. III.
The initial coordinates of the MLWF are set at the center between two [Pd(dmit)$_2$] monomers to 
generate a one-band model (so-called dimer model)~\cite{KinoFukuyama_dimer96}.

The interactions were evaluated by the constrained 
random-phase approximation (cRPA)~\cite{Aryasetiawan_PRB2004} method using \texttt{RESPACK}. 
The energy cutoff for the dielectric function was set to be 3 Ry. 
The interaction terms are given as follows:
\begin{align}
W(\vec{R}_{1},\vec{R}_{2},\vec{R}_{3},\vec{R}_{4})=
\langle\phi_{\vec{R}_1}\phi_{\vec{R}_2}| H_W|\phi_{\vec{R}_3}\phi_{\vec{R}_4}\rangle,
\end{align}
where $H_{W}$ represents the
interaction term of the {\it ab initio} Hamiltonians.
We only treat the two-body interactions, 
such as density-density interactions $U(\vec{R})= W(\vec{0},\vec{0},\vec{R},\vec{R})$
(that is, the on-site and off-site Coulomb interactions)
and direct exchange interactions $J(\vec{R})= W(\vec{0},\vec{R},\vec{R},\vec{0})$,
because the amplitudes of the other terms are negligibly small.

\begin{table*}[t]
  \begin{tabular}{lccccccccccc} \hline
    Cation & Temperature & $t_a$[meV] & $t_b$ [meV]  & $t_c$ [meV] & $U$ [eV] & $V_a$ [eV] & $V_b$ [eV] & $V_c$ [eV] & $J_a$[meV] & $J_b$ [meV]  & $J_c$ [meV] \\ \hline
    Me$_4$P & RT  &59.3 & 44.0 & 30.9&0.883 & 0.449 & 0.465 & 0.413 &3.26&2.54&1.29\\
     & 8~K  &60.6 & 51.8 & 30.8 & 0.793 & 0.414 &0.427 &0.379 & 4.39 & 3.31 & 1.50\\
    Me$_4$As &RT  & 54.9 & 41.5 & 33.8 & 0.864 & 0.426 & 0.442 & 0.394&2.77&2.37	&1.35\\
     & 5~K  & 63.0	& 49.5 & 30.7	& 0.798 & 0.414 & 0.428 & 0.381 & 4.41 & 3.31 & 1.52 \\
    Me$_4$Sb & RT  &49.7 & 36.5 & 38.9& 0.898 & 0.429 & 0.444 & 0.402&2.67&1.75&1.49\\
     & 5~K  & 56.1 & 45.7 & 35.8 & 0.824 & 0.411	& 0.428 & 0.385 & 3.58 & 2.76 & 1.66
    \\ \hline
    EtMe$_3$P & RT &56.1 & 42.0 & 35.0& 0.889 &0.442& 0.460 & 0.411 & 2.88 & 2.02 & 1.48\\
    EtMe$_3$As & RT &53.7 & 40.9 &37.2&  0.889 & 0.436 & 0.456 & 0.409&2.60&1.84&1.45\\
    EtMe$_3$Sb & RT & 48.8& 35.6 & 41.7& 0.906 &  0.427& 0.449 & 0.406&3.21&2.23&1.40\\ 
     & 5~K & 58.3& 45.4 & 40.6& 0.847 & 0.417 & 0.438 & 0.394& 3.30& 2.62& 1.73\\
    & 4~K& 57.1& 44.6 & 40.3& 0.840 & 0.413 & 0.434 & 0.390& 2.23& 1.64& 1.71\\
 \hline
    Et$_2$Me$_2$P & RT &53.4 & 38.5 & 38.6& 0.947 & 0.478 & 0.497 & 0.450&2.51&2.32&1.43\\
    Et$_2$Me$_2$As & RT &50.2 & 36.3 & 38.7& 0.923 &0.422 & 0.467 & 0.448&3.30&2.63&1.73\\
     & 5~K & 55.6 & 43.8 & 36.8& 0.851&	0.431& 0.449& 0.406& 3.60 & 2.71& 1.65\\
    Et$_2$Me$_2$Sb & RT &48.3 & 33.5 & 45.3 & 0.962 &0.461 & 0.485 & 0.443& 2.41& 1.77& 1.58\\ \hline
  \end{tabular}
\caption{List of parameters obtained by the downfolding method in the dimer-model 
(single-orbital extended Hubbard-type Hamiltonian) for $\beta'$-$X$[Pd(dmit)$_{2}$]$_{2}$.
The values of the parameters at RT and those for EtMe$_3$Sb at 4~K
are the same as in Ref.~\onlinecite{Misawa_2020_PRRes}.
The subscripts $a$, $b$, and $c$ represent the directions of the off-site transfer integrals and
interactions, which are illustrated in Fig.~\ref{fig-trans}(d).
}
\label{TransferUV}
\end{table*}

From the above calculations, the following extended single-band Hubbard-type Hamiltonian is obtained:
\begin{align}
&H=
\sum_{ij,\sigma}t_{ij}(c_{i\sigma}^{\dagger}c_{j\sigma}+{\rm h.c.})
+U\sum_{i}n_{i\uparrow}n_{i\downarrow}
+\sum_{ij}V_{ij}N_{i}N_{j} \notag \\
&+\sum_{ij,\sigma\rho}J_{ij}(c_{i\sigma}^{\dagger}c_{j\rho}^{\dagger}c_{i\rho}c_{j\sigma}
+c_{i\sigma}^{\dagger}c_{i\rho}^{\dagger}c_{j\rho}c_{i\sigma}),
\label{Ham}
\end{align}
where $c^{\dagger}_{i\sigma}$ and $c_{i\sigma}$ are the creation and annihilation operators of an electron with spin $\sigma$ in the Wannier orbital localized at the $i$ th dmit dimers.
The number operators are defined as
$n_{i\sigma}=c_{i\sigma}^{\dagger}c_{i\sigma}$ and
$N_{i}=n_{i\uparrow}+n_{i\downarrow}$.
The obtained model parameters are listed in Table~\ref{TransferUV}.
In this study, we evaluate all the microscopic parameters in the effective Hamiltonians such as the transfer integrals $t_{ij}$,
on-site Coulomb interaction $U$,
off-site Coulomb interaction $V_{ij}$,
and {direct} exchange interactions $J_{ij}$ in an $ab$ $initio$ way.

We note that the obtained Hamiltonians are three dimensional.
Since the dmit salts are quasi-two-dimensional compounds, eliminating the weak three dimensionality is justified, and this greatly reduces the numerical cost of analyzing the Hamiltonians. 
In a previous study~\cite{Nakamura_JPSJ2010}, 
the dimensional downfolding method was proposed to obtain two-dimensional Hamiltonians by eliminating the three dimensionality.
It is shown that dimensional downfolding induces a constant shift $\Delta_{\rm DDF}$ 
in the on-site and off-site Coulomb interactions~\cite{Nakamura_JPSJ2010,Nakamura}.
Based on the previous study, in the analysis of the Hamiltonians, we take a constant shift $\Delta_{\rm DDF}=0.30$ eV, 
which is a value comparable to that obtained for EtMe$_3$Sb salt~\cite{Nakamura}, for all compounds.

\subsection{Analysis of effective Hamiltonians}
To clarify how the differences in the low-energy effective Hamiltonians affect the physical quantities, 
we analyze the ground states of the low-energy effective Hamiltonians for nine dmit salts with RT structures 
and five dmit salts with LT structures.
We take $t_{ij}$, $V_{ij}$, and $J_{ij}$ up to the next-nearest neighbor.
To analyze the ground states of the target compounds,
it is desirable to use the crystal structure at the lowest temperature. 
However, to obtain information on how structural differences affect electronic properties, 
it is useful to analyze the effective Hamiltonians for the RT structure 
even if they are hypothetical structures.
For example, as we detail below, shrinking of the lattice parameters is essential for realizing quantum spin liquid behavior in EtMe$_3$Sb salt.

To obtain the ground states, 
we use the locally optimal block conjugate gradient (LOBCG) method~\cite{LOBCG}, 
which is implemented in H$\Phi$~\cite{hphi}.
In this study, we employ a $4\times 4$ system size with twisted boundary conditions.
In the strongly correlated region, the finite-size effects are expected to be small.
However, lowering the temperature increases the system's weakly correlated region and the finite-size effects become significant.
As we show later, we find that the amplitudes of the spin structure factors depend 
on the boundary conditions for Me$_4$P salt in a LT structure,
which has a relatively small $U/t_{a}$.
To reduce the finite-size effects, we perform a
boundary condition average~\cite{twisted_boundary} by introducing the flux $\vec{\phi}$ as follows:
\begin{align}
&c_{i\sigma}^{\dagger}\rightarrow c_{i\sigma}^{\dagger}e^{i\vec{\phi}\cdot\vec{r}_{i}}, \\
&c_{i\sigma}\rightarrow c_{i\sigma}e^{-i\vec{\phi}\cdot\vec{r}_{i}}, \\
&\vec{\phi} = (\phi_x,\phi_y),\
\end{align}
where $\vec{r}_{i}$ is the position vector.
From this flux insertion, the hopping terms and the pair-hopping terms
change as follows:
\begin{align}
c_{i\sigma}^{\dagger}c_{j\sigma}&\rightarrow c_{i\sigma}^{\dagger}c_{j\sigma}e^{i\vec{\phi}\cdot(\vec{r}_{i}-\vec{r}_{j})}, \\
c_{i\sigma}^{\dagger}c_{j\rho}c_{i\rho}^{\dagger}c_{j\rho}
&\rightarrow c_{i\sigma}^{\dagger}c_{j\rho}c_{i\rho}^{\dagger}c_{j\rho} e^{2i\vec{\phi}\cdot(\vec{r}_{i}-\vec{r}_{j})}.
\end{align}
We note that the flux insertion does not change 
the Coulomb interactions.

In a previous study~\cite{twisted_boundary}, it was shown that the finite-size effects are significantly reduced by averaging 
the physical quantities over $\vec{\phi}$.
In our calculations,
we take $-\pi\leq\phi_x<\pi$ and $0\leq\phi_y\leq\pi$ with a division width of $\pi/8$.
In total, we calculate $N_{\phi}=16\times9=144$ boundary conditions.
The boundary averaged spin structure factors are defined as
\begin{align}
&S(\vec{q},\vec{\phi})=\frac{1}{N_{s}}\sum_{i,j}\langle\vec{S}_{i}\cdot\vec{S}_{j}\rangle_{\vec{\phi}} e^{i\vec{q}(\vec{r}_{i}-\vec{r}_{j})}, \\
&S_{\rm ave}({\vec{q}})=\frac{1}{N_{\phi}}\sum_{\phi_x,\phi_x}S(\vec{q},\vec{\phi}), \\
&S({\vec{q}_{\rm peak}})= \max_{\vec{q}}[{S_{\rm ave}({\vec{q}})}],
\end{align}
where $\langle\cdots\rangle_{\vec{\phi}}$ means the expectation value of the ground state for the given boundary condition $\vec{\phi}$.
$N_{\rm s}$ represents the number of sites.
The uncertainties in the boundary average are 
estimated as the standard errors $\sigma/\sqrt{N_{\rm \phi}}$,
where $\sigma$ is the standard deviation in the boundary average.

We also calculate the charge gap $\Delta_{\rm c}$ as follows:
\begin{align}
&\mu^{+}(\vec{\phi}) = \frac{E_0(N_{\rm s}+2,\vec{\phi})-E_0(N_{\rm s},\vec{\phi})}{2}, \\
&\mu^{-}(\vec{\phi}) = \frac{E_0(N_{\rm s},\vec{\phi})-E_0(N_{\rm s}-2,\vec{\phi})}{2}, \\
&\mu^{+}_{\rm min} = \min_{\vec{\phi}}\mu^{+}(\vec{\phi}), \\
&\mu^{-}_{\rm max} = \max_{\vec{\phi}}\mu^{-}(\vec{\phi}), \\
&\Delta_{\rm c} =\max{\{\mu^{+}_{\rm min}-\mu^{-}_{\rm max},0\}},
\end{align}
where $E_0(N,\vec{\phi})$ is the ground-state energy of a system with $N$ electrons and flux $\vec{\phi}$.
The previous study showed that this definition of the charge gap can distinguish metallic and insulating phases well even for small system sizes~\cite{twisted_boundary}.
It was also found that the absolute values of the charge gaps are slightly overestimated due to the finite-size effects.

\begin{figure*}[t] 
\begin{center} 
\includegraphics[width=1.0 \textwidth]{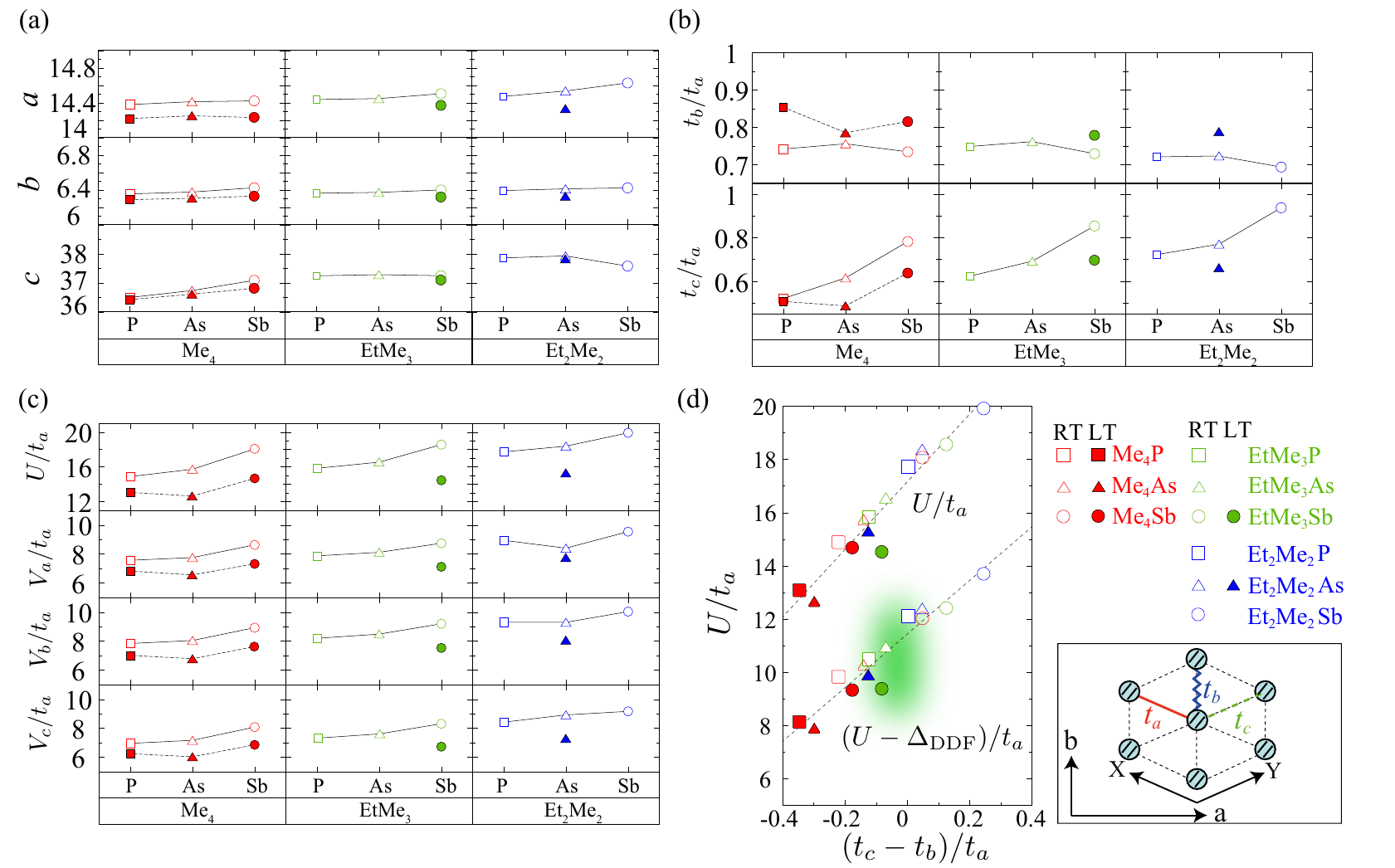}
\vspace{-0.5cm} 
\caption{(a) Lattice constants, (b) transfer integrals, and (c) Coulomb interactions 
for quasi-two-dimensional organic molecules $\beta^\prime$-X[Pd(dmit)$_2$]$_2$ 
(monovalent cations X = Me$_4$Y, EtMe$_3$Y, and Et$_2$Me$_2$Y, where the pnictogen Y is P, As, or Sb). 
The open and closed symbols represent the data obtained using x-ray structures at RT and LT, respectively. (d) Compound dependence of $U/t_a$, $(U-\Delta_{\rm DDF})/t_a$, and $(t_c-t_b)/t_a$. 
The inset shows the schematic of the lattice structure. The closed circles indicate MLWFs and 
each bold line in the closed circles shows a dmit molecule.} 
\label{fig-trans}
\end{center} 
\end{figure*}

\section{Numerical results}\label{Sec:NumericalResults}
In this section, we first show the effective {\it ab initio} 
Hamiltonians for $\beta^\prime$-X[Pd(dmit)$_2$]$_2$ (X represents monovalent cations Me$_4$Y, EtMe$_3$Y, and Et$_2$Me$_2$Y, where Y is a pnictogen, P, As, or Sb).
The values of the parameters such as the transfer integrals and Coulomb interactions for each compound are summarized in Table~\ref{TransferUV}.
Next, we present the numerical results of solving effective Hamiltonians using the exact diagonalization method.

\subsection{Temperature dependence of effective Hamiltonians for $\beta$'-X[Pd(dmit)$_2$]$_2$}
Figures ~\ref{fig-bands}(a) and~\ref{fig-bands}(b) show band dispersions of EtMe$_3$Sb salt for (a) $5$~K and (b) 
RT structures, respectively. 
The solid lines are band structures directly calculated by DFT calculations, 
while the broken lines are those obtained from the transfer integrals between MLFWs defined in Eq.~(\ref{eq:transfer}). 
Note that the two bands crossing the Fermi energy arise due to the presence of the two dimers in the primitive cell.
Focusing on the bands that cross the Fermi energy and the closest bands above and below them, 
we see that the band width increases, i.e., the band crossing the Fermi energy increases about 10\% from RT to the lowest temperatures. 
This increase in the band width is due to the increase in the transfer integrals of $t_a$ and $t_b$ as listed in Table~\ref{TransferUV}, 
and caused by the lattice shrinking with decreasing temperature, as described later.
In Fig.~\ref{fig-bands}(c), the MLFW in EtMe$_3$Sb salt at $5$K is depicted. 
It can be seen that the MLFW is formed as a dimer unit consisting of two [Pd(dmit)$_2$] monomers.

For comparison, the band structures of the LT and RT structures of the Me$_4$P salt 
are shown in Figs.~\ref{fig-bands}(d) and~\ref{fig-bands}(e).
We also find an approximately 8~\% enhancement of the band width of the low-energy band toward low temperatures. 
Unlike other $\beta^\prime$ salt, as the temperature decreases, only $t_b$ increases, 
while $t_a$ and $t_c$ hardly change in Me$_4$P salt.
Lastly, we also show that the MLFW in Me$_4$P salt is formed as a dimer unit, similar to EtMe$_3$Sb salt, as plotted in Fig.~\ref{fig-bands}(f). 

Next, the compound and temperature dependence of the transfer integrals are described.
Figure ~\ref{fig-trans}(a) shows the lattice constants of each compound.
At RT (open symbol), the lattice constants tend to increase in the order of P, As, and Sb for each cation family (Me$_4$Y, EtMe$_3$Y, and Et$_2$Me$_2$Y). 
As the temperature decreases, the lattice generally shrinks, so the overall lattice constant tends to decrease. 
In fact, the lattice constants get smaller.
Figure~\ref{fig-trans}(b) shows the transfer integrals of each compound. 
At RT (open symbols), $t_c/t_a$ tends to increase in the order P, As, and Sb, while $t_b/t_a$ does not greatly depend on the pnictogen.
It can also be seen that $t_b$ does not change much when the cation family 
is changed at the same temperature, while $t_c$ tends to increase in the order Me$_4$Y, EtMe$_3$Y, and Et$_2$Me$_2$Y.
As seen from Table \ref{TransferUV}, $t_a$ and $t_b$ increase, while $t_c$ decreases at low temperatures compared to RT. 
Since the rate of change is larger for $t_b$ than $t_c$, $t_b/t_a$ increases while $t_c/t_a$ decreases. This indicates that the two-dimensionality becomes stronger at low temperature. 
The cation and temperature dependencies of the transfer integrals coincide with the previous studies 
where the transfer integrals are obtained by the extended H\"{u}ckel method\cite{Crystals12_Kato, KUeda_EtMe3Sb2018}. 
It has been reported that the changes in $t_a$ and $t_b$ are mainly due to the distance between the dimers, while the change in $t_c$ is due to the degree of freedom of the atoms in the dimer. 

\begin{figure}[t] 
\begin{center} 
\includegraphics[width=0.48\textwidth]{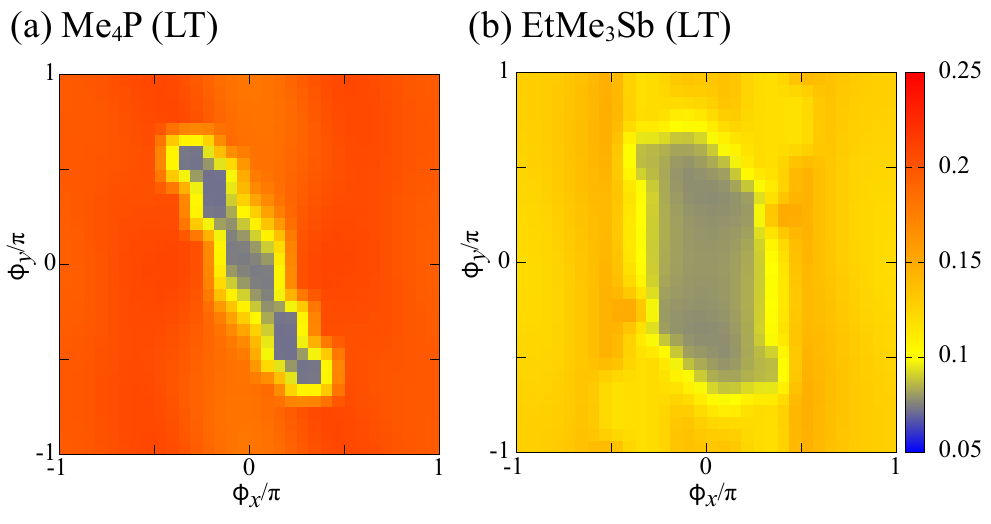}
\end{center} 
\vspace{-0.5cm} 
\caption{
Boundary-condition dependence of the peak value of
the spin structure factors $\left[S(\vec{q}_{\rm peak},\vec{\phi})/N_{\rm s}\right]$
for the effective Hamiltonians of~(a) Me$_4$P with an LT structure and
~(b) EtMe$_3$Sb with an LT structure. 
For Me$_4$P, $\vec{q}_{\rm peak}=(\pi,0)$ for all the boundary conditions.
Although $\vec{q}_{\rm peak}=(\pi,0)$ for EtMe$_3$Sb for most boundary conditions,  
$\vec{q}_{\rm peak}=(\pi,\pi/2)$ in 30 boundary conditions.
As we explained in Sec. II B, we calculate
$-\pi\leq\phi_x<\pi$ and $0\leq\phi_y\leq\pi$ with division width $\pi/8$.
Using the relations $S(\vec{q}_{\rm peak},
\vec{\phi})=S(\vec{q}_{\rm peak},-\vec{\phi})$ and $S(\vec{q}_{\rm peak},(\phi_x=\pi,\phi_y))=S(\vec{q}_{\rm peak},(\phi_x=-\pi,\phi_y))$,
we plot the peak values in the whole Brillouin zone.}
\label{TBC_Sq}
\end{figure}

In molecular solids, the Coulomb interactions tend to be relatively large because the molecules are bound by van der Waals interactions and the transfer integrals become small. 
It is worth noting here that the electron cloud spreads into the molecule. As a result, the on-site Coulomb interaction on the molecule is relaxed. On the other hand, long-range Coulomb interactions between molecules are less affected. Therefore, the long-range Coulomb interaction remains large in organic compounds and needs to be taken into account. From this point of view, we focus on the compounds and structure dependence of the screened Coulomb interactions.
In Fig.~\ref{fig-trans}(c), the effective Coulomb interactions ($U/t_a,~V_a/t_a,~V_b/t_a,~V_c/t_a$) are shown. 
At RT (open symbols), the effective Coulomb interactions gradually increase in the order P, As, and Sb, as well as the transition integral $t_c/t_a$. 
The cation family dependence of the effective Coulomb interaction shows the same trend as $t_c/t_a$, i.e., the effective Coulomb interaction increases in the order Me$_4$Y, EtMe$_3$Y, and Et$_2$Me$_2$Y.
As seen from Table~\ref{TransferUV}, $U$ is almost the same for each cation family. 
Thus the difference of $U/t_a$ in the same cation family is mainly due to the difference in $t_a$. Although $V$ varies in each cation family, the ratio of change for $t_a$ is larger, so $V/t_a$ gives the same trend as $U/t_a$.
At low temperatures, the band width is broadened due to the increase of $t_a$.
As a result, the screened effect becomes large and the screened Coulomb interactions $U,~V_a,~V_b$, and $V_c$ are reduced, as shown in Table \ref{TransferUV}.The trend becomes more pronounced for the effective interactions due to the increase of $t_a$. As seen from Fig.~\ref{fig-trans}(c), the effective Coulomb interactions at low temperatures are weakened by about 20\% compared to those 
at RT. 

\begin{figure}[t] 
\begin{center} 
\includegraphics[width=0.48\textwidth]{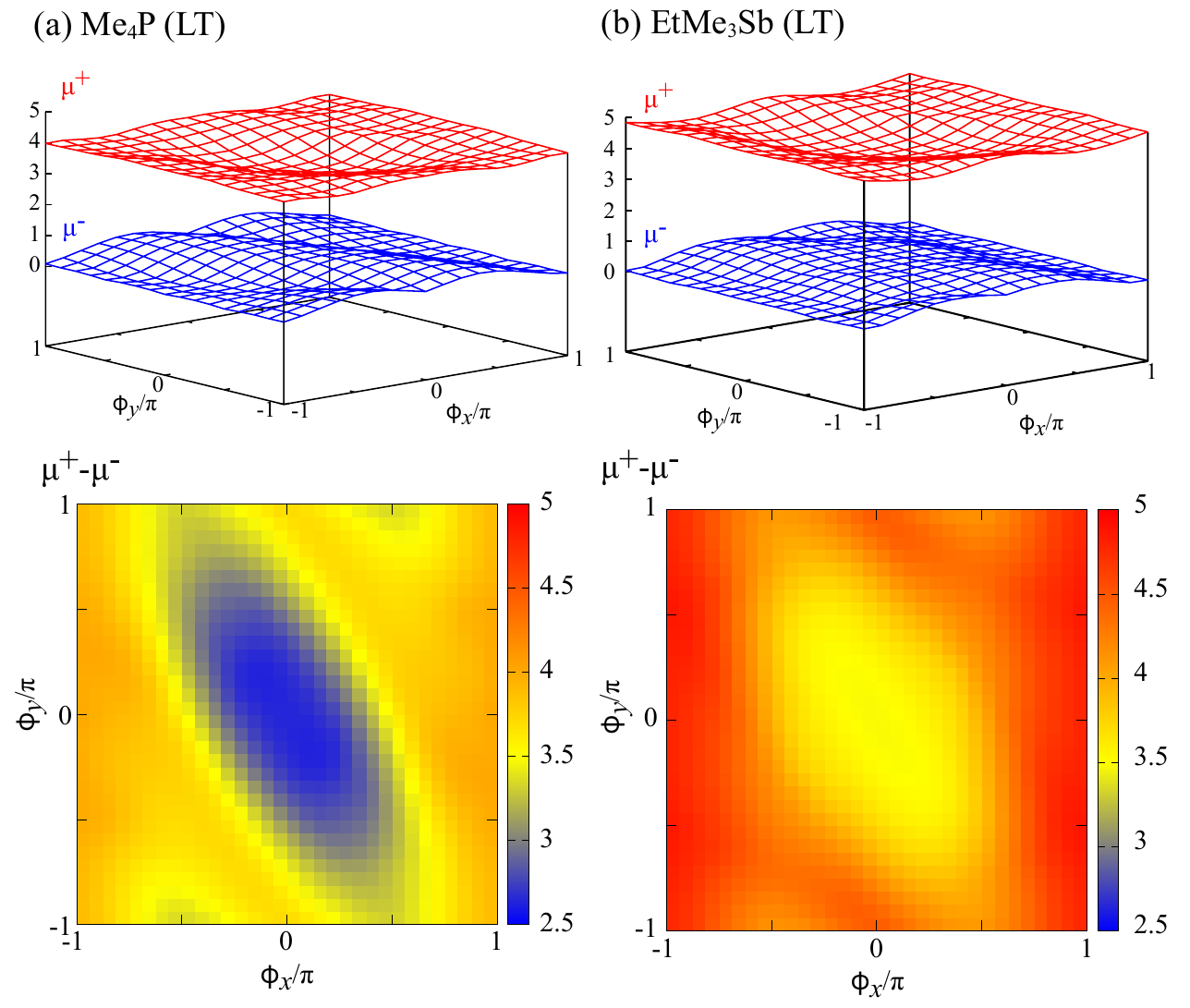}
\end{center} 
\vspace{-0.5cm} 
\caption{Boundary-condition dependence of the chemical potentials
$\mu^{+}(\vec{\phi})$
and $\mu^{-}(\vec{\phi})$
for the effective Hamiltonians of~(a)Me$_4$P salt with an LT structure and
~(b)EtMe$_3$Sb salt with an LT structure.
For clarity, the energy is shifted so that $\min_{\vec{\phi}}\mu^{-}(\vec{\phi})$ is zero.
In the bottom panel, $\mu^{+}(\vec{\phi})-\mu^{-}(\vec{\phi})$ is shown.}
\label{TBC_Gap}
\end{figure}

In a previous study~\cite{Misawa_2020_PRRes}, we showed that the effective on-site Coulomb interaction $U/t_a$ and the anisotropy of the transfer integral $t_c/t_a$ are key parameters for understanding the magnetic ground states in dmit salts. 
In Fig.~\ref{fig-trans}(d), the compound dependencies of $U/t_a$ and $(t_c-t_b)/t_a$ are shown. Here, we adopt $(t_c-t_b)/t_a$, a parameter indicating the magnetic anisotropy, instead of $t_c/t_a$. As mentioned before, $U$ and $t_c$ change in almost the same way, while $t_b$ remains the same among the cation families. 
Therefore, $U/t_a$ tends to be proportional to $(t_c-t_b)/t_a$.
Furthermore, as seen from Table \ref{TransferUV}, $t_a$ is larger and $U$ and $t_c$ are smaller at low temperatures compared to at RT.
Therefore, the trend that $U/t_a$ and $(t_c-t_b)/t_a$ decrease with decreasing temperature is obtained.
We also plot the $(t_c-t_b)/t_a$ dependence of $(U-\Delta_{\rm DDF})/t_a$, where $\Delta_{\rm DDF}=0.30$ is a constant shift parameter to take into account considering the dimensional downfolding effect. It can be seen that the slope becomes a little more gradual, although the trend is basically the same as for $U/t_a$.
As will be detailed in the next section, the magnetic order moment decreases 
in the effective Hamiltonians of materials toward $(t_c-t_b)/t_a \sim 0$ (shaded area), 
and a QSL-like behavior is seen.

Finally, we compare the transfer integrals and Coulomb interactions obtained in our calculations with those calculated in previous studies.
For RT structures, our obtained values of transfer integrals derived from MLWFs reasonably agree with those calculated with a different DFT and fitting methods~\cite{Tsumu_Pd_dmit2_13, Jacko_PRB2013, Seo_2015JPSJ}. 
Similar to the cation dependence found in the RT structures, in the LT structure, $t_b$ is larger than $t_c$ for the salts that show AF order. This difference decreases in the compounds with lower N\'{e}el temperatures and, then, the two transfer energies become comparable around the EtMe$_3$Sb salt, experimentally showing the QSL state~\cite{Kenny_PRM2020}. 
For the magnitude of the Coulomb interactions, the cRPA calculation was performed for the 4~K structure of EtMe$_3$Sb salt by Nakamura $et~al$.~\cite{Nakamura} and the results before dimensional downfolding are in good agreement. In addition, $U/W$ is given by $1.94$ in our calculation for EtMe$_3$Sb salt at 5~K, while, experimentally, it has been reported that $U/W$ is estimated as $2.3$~\cite{Pustogow_NMat_2018}, which is consistent with our results.

\subsection{Analysis of effective Hamiltonians}
Here, we solve the low-energy effective Hamiltonians
for dmit salts using the exact diagonalization for the $4\times4$ cluster.
As explained in Sec.II~B, we employ a boundary-condition average to reduce the finite-size effects.
As a typical example of boundary-condition
dependence of the spin structure factors,
we show $S(\vec{q}_{\rm peak},\vec{\phi})/N_{\rm s}$
of the effective Hamiltonians for Me$_4$P and EtMe$_3$Sb salts with
LT structures in Fig.~\ref{TBC_Sq}.
For both compounds, we find that the spin structure factors become small around $\vec{\phi}=\vec{0}$, i.e. periodic-periodic boundary conditions.
For Me$_4$P salt, except for a narrow region around $\vec{\phi}=\vec{0}$, the spin structure factors show a clear signature of stripe magnetic ordering $\left[\vec{q}_{\rm peak}=(\pi,0)\right]$.
In contrast to this, $S(\vec{q}_{\rm pear},\vec{\phi})/N_{\rm s}$ 
in EtMe$_3$Sb salt is still small sufficiently far from $\vec{\phi}=\vec{0}$.
This significant suppression of the spin structure factors for all the boundary conditions
indicates the appearance of the quantum spin liquid state in EtMe$_{3}$Sb salt.  

In Fig.~\ref{TBC_Gap}, we show $\mu^{+}(\vec{\phi})$ and $\mu^{-}(\vec{\phi})$ as functions of the flux $\vec{\phi}$ for the effective Hamiltonians for Me$_4$P and EtMe$_3$Sb salts with LT structures.
As explained in Sec.II~B, the minimum value of the indirect gap between $\mu^{+}(\vec{\phi})$  and $\mu^{-}(\vec{\phi})$ is the charge gap $\Delta_{\rm c}$.
Since $\mu^{+}(\vec{\phi})$ and $\mu^{-}(\vec{\phi})$ 
do not overlap, the charge gap is finite and thus both compounds are in insulating phases.
The estimated charge gap is $\Delta_{\rm c}=2.6t_{a}$ for
Me$_4$P salt and $\Delta_{\rm c}=3.3t_{a}$ for EtMe$_3$Sb salt. 
This difference is mainly attributed to the amplitude of the Coulomb interactions. 
Since the Coulomb interactions of Me$_4$P salt~[$(U-\Delta_{\rm DDF})/t_a\sim 8.2$] are smaller than 
that of EtMe$_3$Sb salt~[$(U-\Delta_{\rm DDF})/t_a\sim 9.5$], 
the charge gap becomes small.
In spite of the large Coulomb interactions, 
in EtMe$_3$Sb salt, the geometrical frustration as well as the 
off-site Coulomb interaction melt the long-range magnetic order 
and induce the quantum spin liquid behavior.

In the bottom panel of Fig.~\ref{TBC_Gap},
we show $\mu^{+}(\vec{\phi})-\mu^{-}(\vec{\phi})$,
which corresponds to the direct gap, i.e., the charge gap for each boundary condition.
For both compounds, we find that the region where the direct gap becomes small is 
roughly consistent with the region where the
magnetic ordered moment becomes small.
Since the ground state of Me$_4$P salt is a magnetically ordered phase,
this correspondence is natural, i.e., a small magnetic ordered moment means a small charge gap. 
However, this correspondence still holds for EtMe$_3$Sb salt, 
where the magnetic order is suppressed.
%Although we do not have definitive answer to explain the correspondence at this stage, 
The correspondence indicates that the gap generation mechanism 
in the quantum spin liquid is still closely related with the development of the spin correlations.

\begin{figure}[t] 
\begin{center} 
\includegraphics[width=0.45\textwidth]{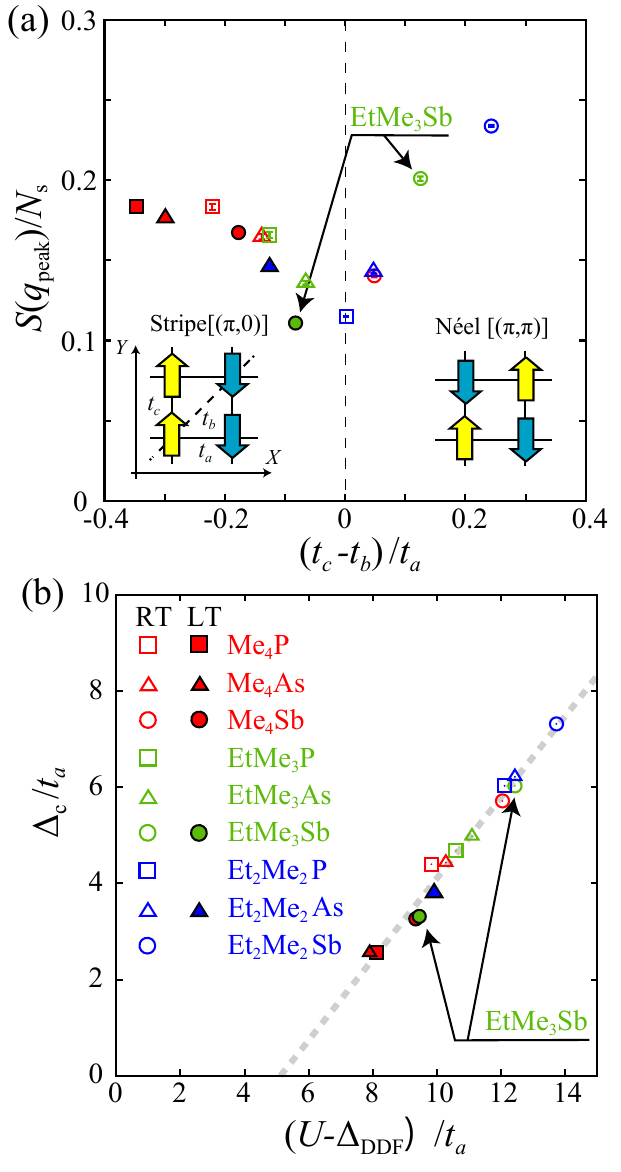}
\end{center} 
\vspace{-0.5cm} 
\caption{(a) Compound dependence of the peak values of the 
boundary averaged spin structure factors $S(\vec{q}_{\rm peak})/N_{\rm s}$ as
a function of $(t_c-t_b)/t_a$.
For $(t_{c}-t_{b})/t_{a}\ll0$, the stripe magnetic ordered phase becomes stable, while
the N\'{e}el-type magnetic ordered phase becomes stable for $(t_{c}-t_{b})/t_{a}\gg0$.
Sandwiched by two magnetic ordered phases,
around $(t_{c}-t_{b})/t_{a}=0$, the spin structure factors are significantly reduced.
The error bars are estimated as the standard errors of the boundary average (see main text).
(b)~Compound dependence of the charge gap as a function of $(U-\Delta_{\rm DDF})/t_{a}$. 
The on-site-Coulomb dependence of the 
charge gap is fitted by the function
$\alpha(U/t_a-U_c/t_a)$, 
where $\alpha=0.84(3)$ and $U_c/t_a=5.1(2)$. The broken line is the result of the 
fitting.}
\label{ED}
\end{figure}

\begin{figure*}[htpb] 
\begin{center} 
\includegraphics[width=0.9\textwidth]{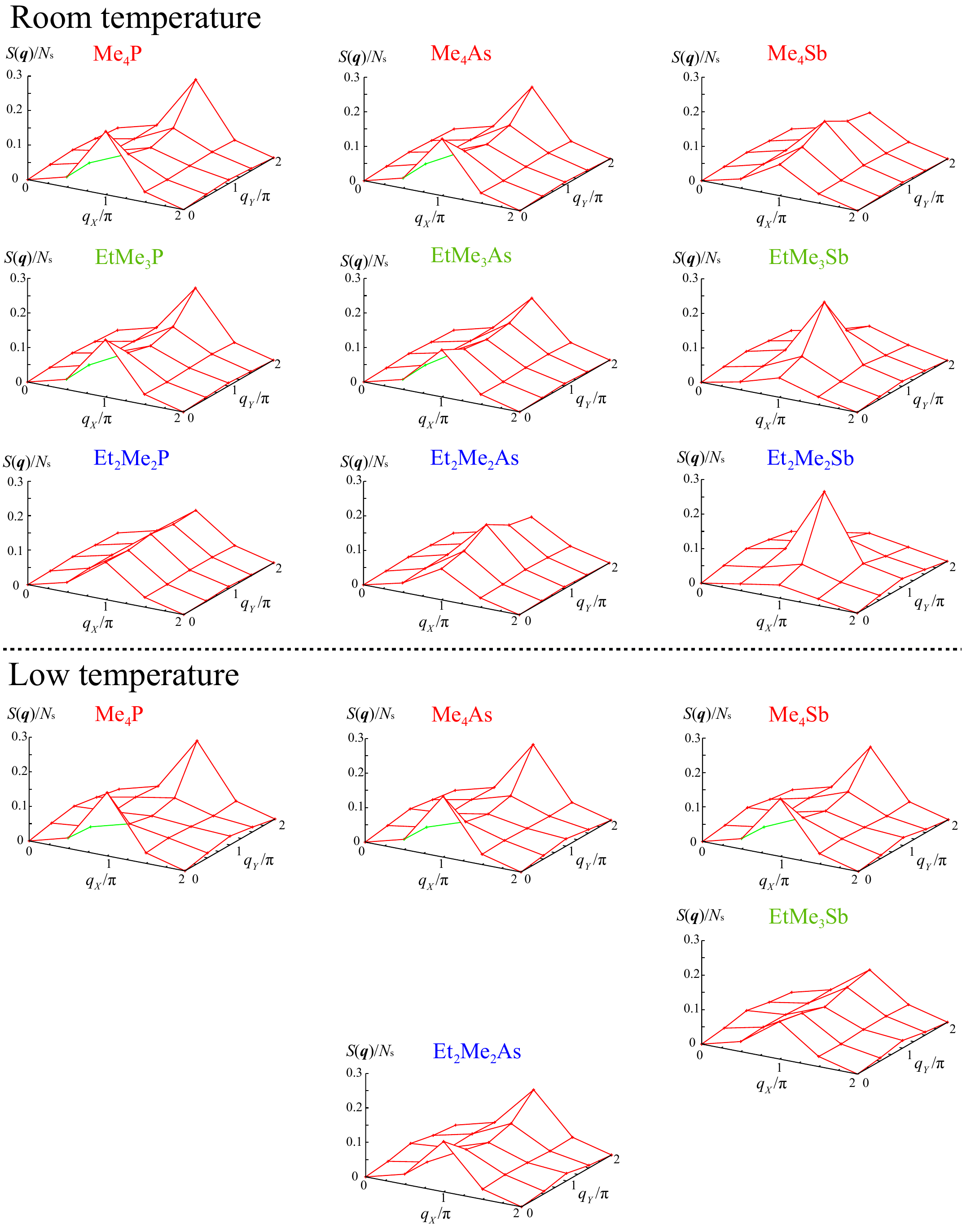}
\end{center} 
\vspace{-0.5cm} 
\caption{~Boundary-averaged spin structure factors for RT structures (upper panel)
and LT structures (lower panel).}
\label{Sq_All}
\end{figure*}

Here, we discuss the compound dependence of the boundary-averaged physical quantities.
First, we calculated the charge structure factors and confirmed that a characteristic peak does not appear and thus all compounds are not in the charge ordered state. Thus, in the following, we only focus on the spin state.
In Fig.\ref{ED}(a),
we show the peak values of 
the boundary averaged spin structure factors [$S({\vec{q}_{\rm peak}})$].
All the spin structure factors in momentum space are shown
in Fig.~\ref{Sq_All}.
Consistent with our previous study,
in EtMe$_3$Sb salt, the spin structure factors are significantly reduced even
when we perform the boundary average.
This reduction is consistent with 
the quantum spin liquid behavior observed in experiments.
Moreover, by performing the boundary average, 
we find that the spin structure factors in EtMe$_3$Sb salt have
weak $q_{y}$ dependence as shown in Fig.~\ref{Sq_All}, i.e.,
implying a larger anisotropy between x and y directions than other compounds,
although the parameters of the transfer integrals and the interactions have no apparent differences in anisotropy compared to those of other compounds.
It is an intriguing future subject to clarify whether this trend survives in larger sized systems.

We examine how the shrinkage of the lattice constants 
associated with lowering temperature 
affects the spin correlations.
As shown in Fig.~\ref{Sq_All},
for all the LT structures, 
we find that stripe-type spin correlations [$q_{\rm peak}=(\pi,0)$]
become dominant since all the compounds are located at $(t_c-t_b)/t_a<0$, while N\'{e}el-type spin correlations [$q_{\rm peak}=(\pi,\pi)$]
become dominant in several RT structures.
For example, in Me$_4$Sb salt, although the 
N\'{e}el-type spin correlation [$\vec{q}_{\rm peak}=(\pi,\pi)$] is dominant in the effective Hamiltonians for the RT structure,
the stripe spin correlation [$\vec{q}_{\rm peak}=(\pi,0)$] becomes dominant in the effective Hamiltonians for the LT structure.
This result indicates that we can observe changes in the spin correlations by lowering the temperature in Me$_4$Sb, by measuring the directions of the spin correlations.

We next examine the compound dependence of the charge gap.
As shown in Fig.~\ref{ED}(b),
for the effective Hamiltonians for LT structures,
we find that the charge gap is significantly reduced because decreasing temperature decreases $U/t_a$. 
For example, in EtMe$_{3}$Sb salt, $\Delta_{c}\sim 6t_{a}\sim 0.29$ eV changes to
$\Delta_{c}\sim 3.3t_{a}\sim 0.18$ eV.
This result indicates that EtMe$_3$Sb salt approaches the boundary of the
metal-insulator transition as the temperature is lowered.
We note that the charge gap at LT ($\Delta_{c}\sim 0.18$ eV) is roughly consistent with 
the experimental value $\Delta_{c}=650{\rm cm}^{-1}\sim0.08$ eV 
estimated by the optical conductivity~\cite{Pustogow_NMat_2018}.
The discrepancy ($\sim 0.1$ eV) can be attributed to the finite-size effects
since the charge gap is expected to be smaller as the system size increases.

\section{Summary and Discussion} \label{Sec:Summary}
In this paper, we have derived low-energy effective Hamiltonians 
for nine dmit salts with RT structures and five dmit salts with LT structures.
We have found that both $U/t_a$ and $(t_c-t_b)/t_a$ decrease 
with decreasing temperature. 
We have obtained the ground states of 
the effective Hamiltonians using the exact diagonalization method with a boundary condition average,
which reduces the finite size effects. 
The results show that the charge gap remains finite in all the materials,
although it decreases with decreasing temperature due to the decrease in $U/t_a$.
By calculating the spin structure factors,
we have found that stripe-type spin correlations appear
in all the effective Hamiltonians for LT structures 
while some of the effective Hamiltonians
for the RT structure shows N\'{e}el-type spin correlations. 
The peak values of the spin structure factors are reduced around
EtMe$_3$Sb salt for LT structures, because $(t_c-t_b)/t_a\sim0$ is satisfied
in the effective model of EtMe$_3$Sb salt.
This result indicates that geometrical frustration
in the inter chain couplings is a key to inducing the quantum 
spin liquid behavior observed in EtMe$_3$Sb salt.
We note that it has been proposed that inter-chain magnetic 
frustration play an important role inducing QSL in the frustrated Hubbard model~\cite{1D_Hab_Valenti} and the 
Heisenberg model~\cite{Yunoki_PRB2006,Kenny_PRM2020, Kenny_PRM2021}. 
Further theoretical calculations to clarify the differences and 
common features between the QSL in the $ab$ $initio$ Hamiltonians and 
that in the simplified Hubbard/Heisenberg model remain as an intriguing subject of study but are beyond the scope of this paper.
This issue will be studied and presented elsewhere~\cite{Ido}.

Here, we comment on the validity of the dimer model description (i.e., single-orbital description) of the dmit salts. 
A recent nuclear magnetic resonance (NMR) experiment~\cite{Fujiyama_PRL2019} showed that 
spin polarizations occur within dimers in Me$_4$P and Me$_4$Sb salts, where the antiferromagnetic ordered phase is the ground state. 
This result indicates that it is necessary to explicitly consider the inter-dimer degrees of freedom to explain the microscopic structures of antiferromagnetic ordered states.  
However, as we have shown in this study, 
the quantum spin liquid behavior is induced by 
frustration in the magnetic interactions among dimers.
Regarding the reduction of the magnetic ordered moment, it is plausible that the microscopic structure of the spin moment within the dimers does not play an essential role.
The success in reproducing the reduction of the magnetic ordered moment
in EtMe$_3$Sb salt demonstrates the validity of the dimer model descriptions of dmit salts
at least for capturing the essence of the quantum spin liquid behaviors.  
A direct comparison and analysis of the dimer model and multi-orbital models that explicitly include the inter dimer degrees of 
freedom is an intriguing subject left for future studies.

The present analysis shows that the microscopic parameters in the effective Hamiltonians of the dmit salts are sensitive to lattice distortions. 
In fact, the values of the on-site Coulomb interaction 
and the charge gap for the LT structure are found to be reduced by about 20\% compared to those for the RT structures.
The amplitude of the geometrical frustration characterized by $(t_c-t_b)/t_a$ is also greatly affected by the changes in the lattice constants.
By utilizing this flexibility, it may be possible to tune the ground states of dmit salts other than EtMe$_3$Sb to become quantum spin liquids by controlling the lattice constants via hydrostatic/uniaxial pressure.
Using $ab$ $initio$ calculations, it is possible to analyze the pressure effects and predict a way to realize a quantum spin liquid.
It is noted that Neel ordered state has not been observed in Et$_2$Me$_2$Sb salt, although 
it has the largest positive $(t_c-t_b)/t_a$ at room temperature. This is because this salt 
undergoes the structure phase transition with the charge ordering. 
%In this material, the charge ordered (CO) state appears but is not stabilized in our calculation. 
However, we do not find the clear signature of the experimentally observed charge ordered (CO) in the effective Hamiltonians. 
%In this case, 
This result indicates that the more complex effects such as more distant long-range Coulomb interactions and electron-phonon couplings should be taken into account\cite{Seo_2015JPSJ} for reproducing the experimentally observed CO state. 
Nevertheless, if we can use the pressure-induced changes in the structures, %make good use of
it might be possible to realize the Neel ordered state at low temperatures 
instead of the CO state. 
If it is possible, we may be able to study the crossover of spin states, i.e., crossover from the
Neel state to the stripe state.
%without having to consider such complicated effects.}
Detailed studies of the pressure effects for stabilizing the quantum spin liquid and possible superconducting phases under pressure are left as important challenges to be addressed.

\begin{acknowledgments}
We wish to thank K. Ido, M. Imada, S. Fujiyama and R. Kato for their helpful contributions. A part of the calculations was done using the Supercomputer Center, the Institute for Solid State Physics, the University of Tokyo.
KY and TM were supported by Building of Consortia for the Development of Human Resources in Science and Technology, MEXT, Japan.
This work was also supported by a Grant-in-Aid for Scientific Research
No.~21H01041 from the Ministry of Education, Culture, Sports, Science and Technology, Japan.  
This work was supported by MEXT as ``Program for Promoting Researches on the Supercomputer Fugaku'' 
(Basic Science for Emergence and Functionality in Quantum Matter-Innovative Strongly-Correlated Electron Science by Integration of ``Fugaku'' and Frontier Experiments-, Project ID: hp210163).
\end{acknowledgments}

\bibliography{main}

\end{document}